\documentclass{article}
\usepackage{spconf,amsmath,graphicx, amssymb}
\usepackage{siunitx}
\PassOptionsToPackage{hyphens}{url}\usepackage{hyperref}
\hypersetup{
colorlinks=true,
citecolor=blue,
linkcolor=black,
linkbordercolor={1 1 1},
citebordercolor={1 1 1},
}
\usepackage[capitalize]{cleveref}
\usepackage[sort]{cite}
\crefname{equation}{}{}
\Crefname{equation}{}{}
\crefname{figure}{Figure}{Figures}
\Crefname{figure}{Figure}{Figures}

\usepackage{caption}
\usepackage{algpseudocode}
\usepackage{algorithm}
\usepackage{mathtools}
\usepackage{physics}
\algrenewcommand\algorithmicindent{0.8em}
\usepackage{multicol}
\usepackage{siunitx}
\usepackage{letltxmacro}
\LetLtxMacro{\oldcite}{\cite}
\renewcommand{\cite}[1]{\mbox{\oldcite{#1}}}

\usepackage{booktabs,ragged2e}
\usepackage{multirow}

\usepackage[flushleft]{threeparttable}

 \def\baselinestretch{0.99}
\title{Efficient Training Data Generation for Phase-Based DOA Estimation}
\name{Fabian Hübner, Wolfgang Mack, Emanu\"el A. P. Habets }
\address{International Audio Laboratories Erlangen,\sthanks{A joint institution of the Friedrich-Alexander-University Erlangen-Nürnberg (FAU) and Fraunhofer Institute for Integrated Circuits (IIS).} Am Wolfsmantel 33 91058 Erlangen, Germany \\
\{wolfgang.mack,emanuel.habets\}@audiolabs-erlangen.de \\
huebner.fa@gmx.de
}
\begin{document}
%\ninept
%
\maketitle

\begin{abstract}
Deep learning (DL) based direction of arrival (DOA) estimation is an active research topic and currently represents the state-of-the-art. Usually, DL-based DOA estimators are trained with recorded data or computationally expensive generated data. Both data types require significant storage and excessive time to, respectively, record or generate. We propose a low complexity online data generation method to train DL models with a phase-based feature input. The data generation method models the phases of the microphone signals in the frequency domain by employing a deterministic model for the direct path and a statistical model for the late reverberation of the room transfer function. By an evaluation using data from measured room impulse responses, we demonstrate that a model trained with the proposed training data generation method performs comparably to models trained with data generated based on the source-image method.
\end{abstract}
\section{Introduction}
\label{sec:intro}
Sound source localization is a crucial task in array signal processing that is used in applications like sound source separation~\cite{Nikunen2014}, speech recognition~\cite{Tsujikawa2018}, camera surveillance~\cite{Chen2013}, and robot audition~\cite{Loellmann2017}. A special case of source localization is direction of arrival (DOA) estimation, which aims at determining the angular position of a source relative to a sensor array. 
DOA estimation methods can be categorized into classical model-based methods and data-driven methods, which are prevalently implemented using deep neural networks (DNN). 

Popular classical methods include (i) subspace-based methods such as MUSIC \cite{1143830, 4392978} and ESPRIT \cite{32276}, (ii) time difference of arrival (TDOA) based methods \cite{8283703},  (iii) methods based on the  steered response power (SRP) such as SRP-PHAT \cite{DiBiase2000AHL} and (iv) statistical methods such as maximum likelihood (ML) \cite{57542}.

Deep learning (DL) based localization techniques are an active topic in the research community and have recently provided state-of-the-art results~\cite{Chakrabarty2017, Chakrabarty2019}. They can be divided into regression methods, which estimate a continuous quantity, and classification methods, which aim to predict a discrete class label for the DOA ~\cite{8937277}.
Most of the DL methods include a feature extraction step rather than using the raw microphone signals. Popular features include (i) the eigendecomposition of the spatial covariance matrix \cite{7471706} (similar to MUSIC), (ii) generalized cross-correlation (GCC) based features  \cite{7178484, 8461267, 8462024, 7738817}, (iii) modal coherence \cite{8936994}, (iv) the Ambisonics intensity vectors \cite{8521403}, (v) phase and magnitude spectra \cite{8553182} and (vi) phase spectra \cite{Chakrabarty2017, Chakrabarty2019}. Many of the features are phase-based as motivated by physical models and classical DOA estimators
~\cite{DiBiase2000AHL}.

The training data generation for DL-based DOA estimators is typically computationally expensive due to costly model-based simulation techniques (e.g., \cite{Habets2008b}) or has specific hardware requirements when the data has to be measured. One way to generate training data for DL-based DOA estimation is by recording sound emitted from a source (e.g., loudspeaker, human) in real acoustic environments \cite{8461267, 8462024}. This approach is time-consuming and for high-quality datasets
a precise ground truth position is essential, which requires expensive measurement equipment.

Another popular method is the convolution of signals (e.g., speech) with room impulse responses (RIRs) that have either been recorded \cite{7738817, 7471706} or simulated based on the source-image method \cite{Allen1979, Chakrabarty2017, Chakrabarty2019, 8521403, 8553182}. The main drawbacks of these data generation methods are excessive time and storage consumption.  These disadvantages get amplified when the simulation time increases due to a growing number of acoustic conditions, the number of microphones, source positions.
Practically, it is a trade-off between cost, time, and storage consumption and the amount of variability of the data set, which is essential to mitigate the risk of overfitting.

We propose an efficient online training data generation method for phase-based DOA estimation. The proposed method is based on a statistical noise model, a deterministic direct-path model for the point source, and a statistical model \cite{Cook1955, Schroeder1962, polack_1992, polack_1993, badeau_2019} for the reverberation. These reverberation models exhibit good modeling capabilities, as shown by their successful application to dereverberation \cite{habets_2010} and automatic speech recognition \cite{5466143}. 

In an evaluation, we train the neural network (NN) from~\cite{Chakrabarty2017} with data provided by the proposed generation method and compare it to the NNs from~ \cite{Chakrabarty2017, Chakrabarty2019} that were trained with data from computationally expensive simulations based on RIRs.

\section{Phase-based DOA estimation}
\label{sec:Fundamentals}

We consider a microphone array with M microphones that is placed in an enclosed space and receives reverberant sound emitted from a single point source. We denote the cartesian coordinates of each microphone $i \in \{1, \:  \ldots,\: M\}$ by $\mathbf{m}_i$ and the source coordinates by $\mathbf{s}$, and denote the discrete frequency domain microphone signal by $Y_i(k)$. Neglecting spectral leakage and the DC-component, we model $Y_i(k)$ by a multiplicative model with additive noise $N_i(k)$, i.e.,
\begin{equation}
\label{eq:y_mult_model_add_noise}
    Y_i(k) = H_i(k) X(k)+ N_i(k),
\end{equation}
where $X(k)$ is a frequency domain source signal, $H_i(k)$ is a microphone dependent room transfer function (RTF), $k \in \{1, \: \ldots , \: K \}$ is the
frequency index and K is the length of the one-sided discrete Fourier transform (DFT). 
The RTF can be decomposed into a direct part, $H_{i, \textrm{dir}}(k)$ and a late reverberant part, $H_{i,\textrm{rev}}(k)$, i.e.,
\begin{equation}
\label{eq:RTF_decomposition}
\begin{aligned}
    H_i(k) &= H_{i,\textrm{dir}}(k) + H_{i,\textrm{rev}}(k).
\end{aligned}
\end{equation}

The objective of phase-based DOA estimation is to obtain the angle of arrival $\theta$ of the sound source based on the phase map $\mathbf{\Phi}$ of the microphone signals that is defined as
\begin{equation}
\label{eq:phase_map}
    \mathbf{\Phi} = \begin{bmatrix} \angle{Y}_1(:), & \ldots, &\angle{Y}_M(:) \end{bmatrix} \in \mathbb{R}^{K \times M},
\end{equation}
where we use $\angle$-operator to denote the phase extraction.

A state-of-the-art phase-based DOA estimator that uses a DNN was proposed in \cite{Chakrabarty2017} for single-source localization and adapted to a multi-source scenario in \cite{Chakrabarty2019}.
In \cite{Chakrabarty2017, Chakrabarty2019}, the DOA estimation task is formulated as a classification problem with 37 angular classes ranging from \ang{0} to \ang{180} with a resolution of \ang{5}. The input phase map $\mathbf{\Phi}$ is extracted from a uniform linear microphone array with 4 microphones. The DNN in \cite{Chakrabarty2017}  consists of 3 convolutional layers followed by 3 fully connected layers, as described in \cref{tab:architectures}. In \cite{Chakrabarty2019}, a slightly modified architecture was used.

\begin{table}[t]
\small
\begin{threeparttable}
\setlength\tabcolsep{0pt} % make LaTeX figure out intercolumn spacing
\begin{tabular*}{\columnwidth}{@{\extracolsep{\fill}} c rrccc}
\toprule\
     \multirow{2}{*}{Layer } & Input & Output & Kernel & \multirow{2}{*}{Activation} & \multirow{2}{*}{Dropout} \\
     & Shape & Shape & Size &  \\
     \midrule

    Conv1 & 1x256x4 & 64x255x3 & (2,2) & ReLU & No\\
    Conv2 & 64x255x3 & 64x254x2 & (2,2) & ReLU & No\\
    Conv3 & 64x254x2 & 64x253x1 & (2,2) & ReLU & Yes\\
    \midrule  
    Linear1 & 16192 & 512 & --- & ReLU & Yes  \\
    Linear2 &   512 & 512 & --- & ReLU & Yes  \\
    Linear3 &   512 & 37  & --- & Softmax & No   \\
%\addlinespace
\bottomrule
\end{tabular*}
\caption{Network architecture according to \cite{Chakrabarty2017}.}
\label{tab:architectures}
%\vspace{-1em}
\end{threeparttable}
\end{table}

The training data generation in \cite{Chakrabarty2017} and \cite{Chakrabarty2019} is based on RIRs that are simulated for different room geometries and microphone positions using the source-image method~\cite{Allen1979}. The RIRs are convolved with noise source signals, and spatially white microphone noise is added.
The main drawback of this data generation approach is the high computational cost, which is due to (i) the RIR simulation and (ii) the convolutions with long filters. This makes online training unpractical and therefore requires memory to store the generated training data. As the data has to be generated for a specific microphone array geometry, adaptations in the geometry require to repeat the data generation process, which makes the method unsuited for fast prototyping.
\section{Proposed Data Generation Method}
\label{sec:proposed}
\subsection{Signal Model}
To enable online training data generation for arbitrary microphone array geometries, we propose a RIR and convolution free data generation method by modeling the individual components of~\cref{eq:y_mult_model_add_noise}.
We model the source signal  ${X}(k)$  and the additive noise signals ${N}_i(k)$ by zero-mean, circular symmetric, complex Gaussian processes, where we assume statistical independence in the frequency domain. In principle, other application-specific distributions may be incorporated here.
To simplify notation, we consider the source signal ${X}(k)$ to have unit variance and denote the variance of the additive noise signals  ${N}_i(k)$ by $\sigma_N^2$.

We model ${H}_i(k)$ by a deterministic direct path model and a stochastic reverberation model. 
The direct part is modeled as a complex exponential, i.e.,
\begin{equation}
\label{eq:RTF_direct}
    H_{i,\textrm{dir}}(k) = e^{-j \phi_{i,\textrm{dir}}(k)}
\end{equation}
with $j := \sqrt{-1}$ and a microphone dependent phase term $\phi_{i,\textrm{dir}}(k)$, that is given by 
\begin{equation}
\label{eq:RTF_direct_phase}
       \phi_{i,\textrm{dir}}(k) = \frac{||\mathbf{m}_i - \mathbf{s}||_2}{c }   \frac{ \pi f_s k }{K},
\end{equation}
where $c$ denotes the speed of sound, $f_s$ the sampling frequency and $||\cdot||_2$ the $\ell^2$-norm. Assuming the center of the microphone array at $\begin{bmatrix} 0 &  0 & 0 \end{bmatrix}^T$, the source position $\mathbf{s}$ is calculated according to
\begin{equation}
\label{eq:source_positon}
    \mathbf{s} = \begin{bmatrix} r \cos(\theta) & r \sin(\theta) & 0\end{bmatrix}^T,
\end{equation}
where $r$ is the source-microphone distance.

The reverberant part of the RTF $H_{i,\textrm{rev}}(k)$ is considered as a diffuse, isotrophic sound field and is modeled by a zero-mean, circular symmetric, complex Gaussian process \cite{badeau_2019,SCHULTZ197117}. We assume statistical independence of the frequency bins and incorporate spatial correlation by the covariance matrices $\mathbf{\Sigma}_{H} \in 
\mathbb{R}^{M \times M}$, given by
\begin{equation}
\label{eq:covariance}
  \mathbf{\Sigma}_{H}(k) = \sigma_{R}^2 \mathbf{ \Gamma }_{H}(k),
\end{equation}
where $\sigma_R^2$ denotes the reverberation variance and the entries of the spatial coherence matrices $\mathbf{\Gamma}_{H}(k)$ are computed according to Cook's formula~\cite{Cook1955},~i.e.
\begin{equation}
\label{eq:Cook}
\begin{aligned}[b]
    \mathbf{\Gamma}_{H}(k)_{i, j} & := \frac{\mathbb{E}\{H_{i,\textrm{rev}}(k) H_{j,\textrm{rev}}^*(k)\}}{\sqrt{\mathbb{E}\{|H_{i,\textrm{rev}}(k)|^2\}\mathbb{E}\{|H_{j,\textrm{rev}}(k)|^2\}} } \\
    & = \text{sinc}\left( \frac{  ||\mathbf{m}_i - \mathbf{m}_j||_2}{c} \frac{\pi f_s k}{K}\right)
   \end{aligned},
\end{equation}
where $\text{sinc}(x) := \frac{\text{sin}(x)}{x}  \text{ if } x \neq 0; \text{else } 1$; and $H_{j,\textrm{rev}}^*(k)$ denotes the complex conjugate of $H_{j,\textrm{rev}}(k)$. 
The variances $\sigma_R^2$ and $\sigma_N^2$ are related to the decibel domain signal-to-noise ratio $\textrm{SNR}_\textrm{dB} $ and the direct-to-reverberation ratio $\textrm{DRR}_\textrm{dB}$ by
\begin{equation}
\label{eq:SNR_DRR}
    \sigma_{R}^2 = 10^{-\frac{\textrm{DRR}_\textrm{dB}}{10}}\; \text{and}\; \sigma_{N}^2 = 10^{-\frac{\textrm{SNR}_\textrm{dB} }{10}}.
\end{equation}

\subsection{Algorithm}
\label{sec:algorithm}
Based on the previously defined model, the proposed method generates data samples by Monte Carlo simulation. As the problem is formulated as a classification task, first a class label $\theta$ is sampled from a discrete uniform distribution and the parameters $r$, $\textrm{SNR}_\textrm{dB} $ and $\textrm{DRR}_\textrm{dB} $ are sampled from independent continuous uniform distributions. In principle, other distributions are possible, e.g., a distance-dependent DRR distribution, but that was not considered in the current framework. We then calculate the variances $\sigma_R
^2$ and $\sigma_N^2$ according to \cref{eq:SNR_DRR} and the source position according to \cref{eq:source_positon}.

For each set of parameters, the samples are generated according to \cref{algo:rtf_generation,algo:data_generation}, where we use the symbol $\leftarrow$ to denote a sampling process and denote a zero-mean, circular symmetric, complex Gaussian process by $\mathcal{N_C}(\diamondsuit, \boxdot)$, where $\diamondsuit$ and $\boxdot$ are placeholders for the mean and (co)variance parameters, respectively. 

The sample generation is a two-step procedure. In the \cref{algo:rtf_generation} the RTFs $H_i(k)$ are created by calculating the direct part according to \cref{eq:RTF_direct,eq:RTF_direct_phase} and creating correlated reverberation samples according to \cref{eq:covariance,eq:Cook}. \Cref{algo:data_generation} generates samples of the source signal $X_i(k)$ and the additive noise signals $N_i(k)$ and composes the microphone signals $Y_i(k)$ according to \cref{eq:y_mult_model_add_noise}. The algorithm finishes with the feature extraction according to \cref{eq:phase_map}. In practice, the algorithmic steps can be implemented efficiently in vectorized form.
\begin{algorithm}[t]
\small
\caption{RTF generation}
\begin{algorithmic}[1]
\label{algo:rtf_generation}
\Function{gen\_rtf}{$\sigma_R^2, \mathbf{s}, \mathbf{m_1}, \ldots, \mathbf{m_M}$}
        \State  $H_{(:),\textrm{rev}}(k) \ \leftarrow \mathcal{N_C}(\mathbf{0},\: \sigma_R^2 \mathbf{\Gamma}_{H}(k)) \; \forall k$ \Comment{\cref{eq:covariance,eq:Cook}}
        \For{i=1 to M}
            \State calculate $H_{i,\textrm{dir}}(k) \; \forall k$ \Comment{\cref{eq:RTF_direct,eq:RTF_direct_phase}}
            \State $H_{i}(k)  = H_{i,\textrm{dir}}(k) + H_{i,\textrm{rev}}(k) \; \forall k$ \Comment{\cref{eq:RTF_decomposition}}
      \EndFor
      \State \Return $H_{(:)}(:)$
\EndFunction

\end{algorithmic}
\end{algorithm}
\vspace{-1em}

\begin{algorithm}[t]
\small
\caption{Sample generation}
\begin{algorithmic}[1]
\label{algo:data_generation}

\Function{gen\_sample}{$\sigma_R^2, \sigma_N^2, \mathbf{s}, \mathbf{m_1}, \ldots, \mathbf{m_M}$}
    \State $H_{(:)}(:) = $ GEN\_RTF($\sigma_R^2, \mathbf{s}, \mathbf{m_1}, \ldots, \mathbf{m_M}$)  \Comment{\cref{algo:rtf_generation}}
    \State  $X(k) \leftarrow \mathcal{N_C}(0,\;1) \; \forall k$
    \For{i=1 to M}
            \State $N_i(k) \leftarrow \mathcal{N_C}(0,\: \sigma_{N}^2) \; \forall k$
            \State $Y_i(k) = X(k) H_i(k) + N_i(k) \;\forall k $ \Comment{\cref{eq:y_mult_model_add_noise}}
\EndFor
    \State calculate phase map $\mathbf{\Phi}$ \Comment{\cref{eq:phase_map}}
    \State \Return $\mathbf{\Phi}$
\EndFunction
\end{algorithmic}
%\cmidrule{}
\end{algorithm}
%\vspace{-0.5cm}
\begin{figure*}[!ht]
\begin{minipage}[b]{1.0\linewidth}
  \centering
  \centerline{\includegraphics{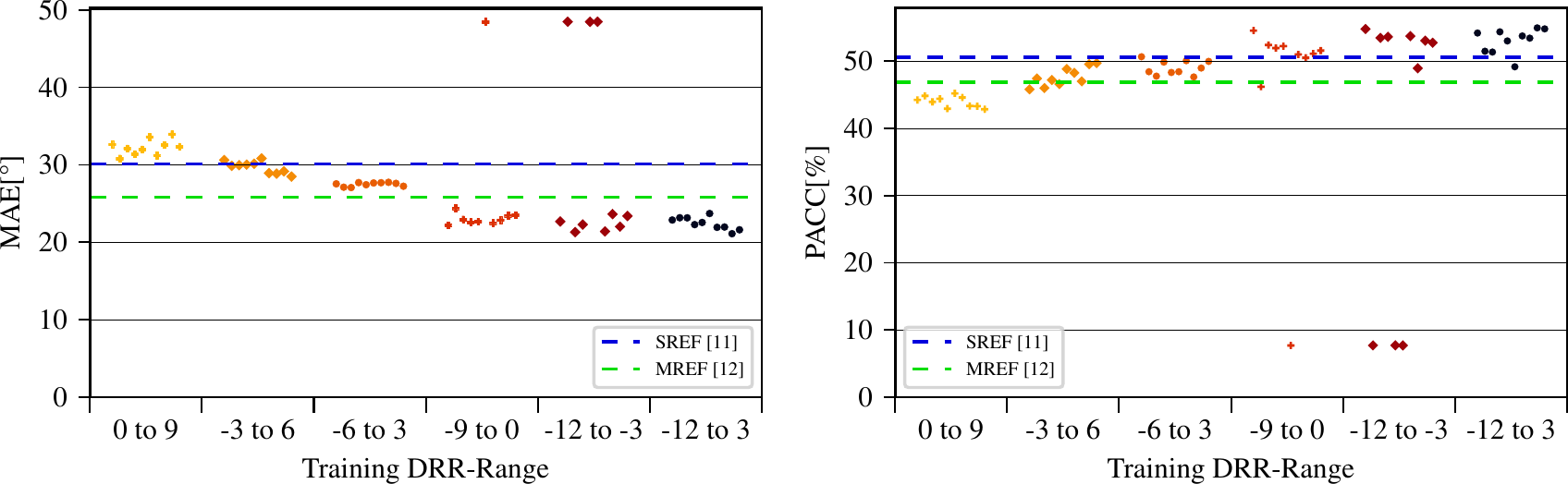}}
\end{minipage}
  \caption{Frame-level performance on the test set for different training DRR-Ranges: For each parameter setting, we trained 10~networks with the proposed data generation method using  different random number generator seeds.}
  \label{fig:drr_performance}
\end{figure*}

\section{Datasets}
\label{sec:datasets}
As in \cite{Chakrabarty2017, Chakrabarty2019, ChakrabartyGit}, we consider a uniform linear array with 4 microphones, an inter-microphone spacing of $0.08$~m, a sampling frequency of $16$~kHz, and a DFT length of~512 for all experiments. The validation set was generated using simulated RIRs with the room parameters reverberation time $T_{60}$ and room dimensions $dim$ given in the lower part of \Cref{tab:datasets}. The RIRs from the validation set are convolved with noise sources. The test set is generated using measured RIRs from~\cite{6954309} (4~central microphones of the $\left[8, \ldots, 8 \right]$
~cm configuration). For the test set, the RIRs are convolved with recordings from the Librispeech corpus \cite{7178964}. The training set is generated online according to the proposed algorithm given in \cref{sec:algorithm}. 
For all datasets, we incorporate additive noise and use different source-microphone distances $r$ and different DOAs $\theta$ as given in the upper part of \cref{tab:datasets}, where $\mathcal{U}(\diamondsuit, \boxdot))$ denotes a continuous uniform distribution with the placeholders $\diamondsuit$ and $\boxdot$ for the lower and upper bounds. The validation set and the test set are calculated using the short-time Fourier transform with a Hann window of length 512 and an overlap of 256 samples. In total, the validation set comprises \num{2536156} samples, the test set comprises \num{156000} samples, and the training data consists of 8000 online generated minibatches with a size of~512. 

\begin{table}[t]
\small
\begin{threeparttable}
\setlength\tabcolsep{0pt} % make LaTeX figure out intercolumn spacing
\begin{tabular*}{\columnwidth}{@{\extracolsep{\fill}} c ccc}
\toprule\
Dataset &  Training &  Validation & Test \tnote{1}  \\
     \midrule
$\textrm{SNR}_\textrm{dB} $ & $\mathcal{U}(0, 30)$ & $\mathcal{U}(0, 30)$ & $\mathcal{U}(10, 30)$ \\

$r\: [m]$ & $\mathcal{U}(1, 3)$ & $\{1.2, 2.3\}$ & $\{1, 2\}$ \\
$\theta\:[^\circ]$ & $\{0, 5, \ldots, 180\}$ & $\{0, 5, \ldots, 180\}$ & $\{0, 15, \ldots, 180\}$ \\
\midrule
\multicolumn{4}{c}{Simulated Rooms for Validation Set} \\
$dim\: [m]$ & [9, 11, 2.7] & [9, 11, 2.7] &  [10, 10, 2.7]\\
$T_{60}\: [s]$ & 0.45 &  0.60 & 0.75 \\
%\addlinespace
\bottomrule
\end{tabular*}
\begin{tablenotes}\footnotesize
\item[1] The room parameters for the test set are given in  \cite{6954309}
\end{tablenotes}
\caption{Dataset parameters}
%\vspace{-1.5em}
\label{tab:datasets}
\end{threeparttable}
\end{table}

\section{Performance evaluation}
\label{sec:evaluation}
We trained the network with the same training parameters as in \cite{Chakrabarty2017} for 8000 minibatches of size 512. For model selection, the mean absolute (MAE)  was calculated after every 100 mini batches based on a 10000 samples sized subset of the validation set, and the model with the lowest MAE was selected. We performed a frame-level evaluation, where the estimate was obtained by picking the class label with the maximum probability, and a block-level evaluation, where the network's output probabilities were averaged first.
For the frame-level evaluation, we use the metrics MAE and the pseudo-accuracy (PACC), which we define to be the prediction accuracy with $5^\circ$ tolerance, i.e., we consider the classification as correct if the distance between the true DOA and the estimate is less than or equal to $5^\circ$ as in~\cite{Chakrabarty2019}.  For the block-level evaluation, we consider a signal block of 50 consecutive frames and define the metrics PACC\textsubscript{50} and MAE\textsubscript{50} to be the PACC and MAE metrics, calculated on the averaged output probabilities of a 50 frame segment. To simplify notation, we abbreviate the baseline from~\cite{Chakrabarty2017} by SREF and the baseline from~\cite{Chakrabarty2019} by MREF. (network weights from~\cite{ChakrabartyGit})

In the first experiment, we demonstrate the influence of the training DRR-Range on the test set performance of the network, as demonstrated in \cref{fig:drr_performance}. For each DRR-parametrization, we trained 10 networks with different random number generator seeds. The performance was evaluated frame-wise. The MAE and PACC performance increases with decreasing DRR until it saturates at the DRR-Range of [-9;~0]. Moreover, a too restrictive DRR-Range can cause the network training to fail: For the [-9;~0] parametrization  1 out of 10 and for the [-12;~ -3] parametrization 3 out of 10 models failed as their MAE performance ($\approx$ $50^\circ$) and PACC performance ($\approx$ 8\%) are on the same scale as an untrained model. The [-12;~ 3] parametrization does not show this behavior, so increasing the upper DRR bound can be a possible mitigation strategy. Disregarding the outliers, the performance is comparable to the baseline methods.

For the second experiment, we select the model with the minimum MAE based on the validation set and evaluate the performance for the different reverberation times $T_{60}$ of the test set separately. For each file, the central 50 frames corresponding to 0.8 s of the speech utterances are selected, and the metrics PACC\textsubscript{50} and MAE\textsubscript{50} are calculated. The results are depicted in \cref{tab:t60}. At the PACC\textsubscript{50} metric, the model trained using the proposed data generation method performs best for a 
 $T_{60}$ of 0.36~s and second-best for the $T_{60}$ values of 0.16~s and 0.61~s. Except for the $T_{60}$ of 0.16, the MAE\textsubscript{50} performance of the model trained using the proposed data generation method is comparable to the baselines. 
 
 Considering both, the MAE/MAE\textsubscript{50} and the PACC/PACC\textsubscript{50} metrics for the frame and block-level evaluation, the overall performance of the model trained using the proposed data generation method is on par with the baselines.
\begin{table}[tb]
\small
\begin{threeparttable}
\setlength\tabcolsep{0pt} % make LaTeX figure out intercolumn spacing
\begin{tabular*}{\columnwidth}{@{\extracolsep{\fill}} c ccc ccc}
\toprule
& \multicolumn{3}{c}{PACC\textsubscript{50}[\%] } & \multicolumn{3}{c}{MAE\textsubscript{50}[$^\circ$]} \\
$T_{60}[s]$ & 0.16 & 0.36  & 0.61  & 0.16 & 0.36  & 0.61 \\ 
 \cmidrule{2-4}  \cmidrule{5-7} 
SREF \cite{Chakrabarty2017} & 87.69 & 89.55 & \textbf{87.12} &  2.25  & 2.41  & \textbf{2.74}\\
MREF \cite{Chakrabarty2019} & \textbf{89.62} & 90.91 & 82.12 &  \textbf{2.07} &   \textbf{1.84} & 4.66\\
Proposed & 87.69 & \textbf{95.00} &  83.85 & 3.73 & 2.16  & 4.70\\
%\addlinespace
\bottomrule
\end{tabular*}
\caption{Block-level performance for different $T_{60}$. The metrics were calculated from the average output propabilities of a 50 frame segment (central 0.8 s)}
\label{tab:t60}
\vspace{-1.0em}
\end{threeparttable}
\end{table}
\section{Conclusion}
\label{sec:conclusion}
We proposed a low complexity model-based training data generation method for phase-based DOA estimation. The proposed method models the microphone phases directly in the frequency domain to avoid computationally costly operations as present in state-of-the-art methods. The low computational complexity of the proposed method allows for online training data generation, which allows faster prototyping, and paves the way for applications with a high data demand such as moving sound sources simulation or large microphone arrays. An evaluation using measured RTFs yielded comparable results for phase-based DOA estimation when using the proposed method and the computationally expensive source-image method for training data generation.

 \def\baselinestretch{0.8775}

%\clearpage
%\vfill\pagebreak
%\newpage
\small
\bibliographystyle{IEEEbib}

\bibliography{refs}

\end{document}